\documentclass[twocolumn,prl,superscriptaddress]{revtex4}

\usepackage{graphicx}
\usepackage{amssymb}
\usepackage{amsfonts}

\DeclareGraphicsExtensions{.eps}

\begin{document}

\preprint{}

\title{Polariton quantum boxes in semiconductor microcavities}
\author{O. El Da\"{\i}f}
\affiliation{Institut de Photonique et d'{\'E}lectronique Quantiques, {\'E}cole Polytechnique F{\'e}d{\'e}rale de Lausanne (EPFL), CH-1015 Lausanne,
Switzerland.}

\author{A. Baas}
\affiliation{Institut de Photonique et d'{\'E}lectronique Quantiques, {\'E}cole Polytechnique F{\'e}d{\'e}rale de Lausanne (EPFL), CH-1015 Lausanne,
Switzerland.}

\author{T. Guillet}
\affiliation{Groupe d'{\'e}tude des semiconducteurs (GES), Universit{\'e} de Montpellier II, Place Eug{\`e}ne Bataillon, F-34095 Montpellier cedex 5,
France}

\author{J.-P. Brantut}
\affiliation{Institut de Photonique et d'{\'E}lectronique Quantiques, {\'E}cole Polytechnique F{\'e}d{\'e}rale de Lausanne (EPFL), CH-1015 Lausanne,
Switzerland.}

\author{R. Idrissi Kaitouni}
\affiliation{Institut de Photonique et d'{\'E}lectronique Quantiques, {\'E}cole Polytechnique F{\'e}d{\'e}rale de Lausanne (EPFL), CH-1015 Lausanne,
Switzerland.}

\author{J.L. Staehli}
\affiliation{Institut de Photonique et d'{\'E}lectronique Quantiques, {\'E}cole Polytechnique F{\'e}d{\'e}rale de Lausanne (EPFL), CH-1015 Lausanne,
Switzerland.}

\author{F. Morier-Genoud}
\affiliation{Institut de Photonique et d'{\'E}lectronique Quantiques, {\'E}cole Polytechnique F{\'e}d{\'e}rale de Lausanne (EPFL), CH-1015 Lausanne,
Switzerland.}

\author{B. Deveaud}
\affiliation{Institut de Photonique et d'{\'E}lectronique Quantiques, {\'E}cole Polytechnique F{\'e}d{\'e}rale de Lausanne (EPFL), CH-1015 Lausanne,
Switzerland.}

\begin{abstract}
We report on the realization of polariton quantum boxes in a semiconductor microcavity under strong coupling regime. The quantum boxes consist of mesas, etched on the top of the spacer of a microcavity,
that confine the cavity photon. For mesas with sizes of the order of a few microns in width and nm in depth,
we observe quantization of the polariton modes in several states, caused by the lateral confinement. We evidence the strong exciton-photon coupling regime
through a typical  anticrossing curve for each quantized level. Moreover the growth technique permits to obtain high-quality samples, and opens the way for the
conception of new optoelectronic devices.

\end{abstract}


\date{\today}

\maketitle

\newpage

Confining semiconductor structures allows the study of various fundamental effects, ranging from the Purcell effect to the full quantum confinement.  Such
confinement is also used for applications in many fields, from optoelectronics to quantum information. Previous works have focused on different aspects: on
the one hand, on the matter part, with the confinement of the excitonic resonances in quantum wells, quantum wires and quantum dots. On the
other hand, environment for the electromagnetic field has been modified by optical confinement in different types of cavities.  Additionally since the middle of
the 90's, low dimensional devices have been realized in the strong coupling regime \cite{Weisbuch92}. Confinement can enhance the interactions, modify the real and imaginary parts of the resonance's energy, or open access to new interaction processes. It is also often considered as a possible way to obtain a condensed phase of bosons in semiconductors \cite{Snoke}, but so far the fermionic nature of excitons has always become dominant upon increasing density. In this sense, polaritons are of great interest as, despite their excitonic content, they have a very small effective mass in comparison to the exciton (thanks to
their photonic component), which theoretically increases their temperature of condensation (above $0.1K$) \cite{kavokin}. The peculiar trap shape of the lower microcavity polariton dispersion curve has motivated several relaxation experiments towards the bottom of this "trap" \cite{Baumberg00,Richard05} but no clear evidence for the formation of spontaneous
coherence formation has been given yet.

0D Polariton confinement can be achieved either through their excitonic or through their photonic component. Recently, evidence for 0D polaritons has been
given with single quantum dots in micropillars \cite{Reithmaier}, photonic nanocavities \cite{Yoshie}, or microdisks \cite{Peter} and for a large number of excitations in
micropillar structures \cite{bloch,obert,dasbach}. Here we consider a novel system under strong coupling regime, where 0D confinement is achieved through
the photonic part of polaritons in high Q cavities. Our original structure contains polariton quantum boxes, constituted by mesas in the spacer layer of a
semiconductor microcavity, allowing to keep the strong coupling regime. Each mesa, by acting on the two degrees of freedom of the photonic component of
the 2D cavity polaritons, does create a localized photonic box in the microcavity.

The main specificity of our technique is that the semiconductor microcavity around the mesas is in no way altered by the creation of the box, contrary to
the case of etched microcavities. This brings about a number of advantages : i) the presence of the 2D cavity restricts the lateral losses for the confined
mode, ii) the number of confined levels in the quantum box is controlled by the height of the mesa, iii) interaction between the 0D and the surrounding 2D
polaritons is possible, iv) the technique also allows to create patterns at will, for example polariton quantum wires \cite{dasbach1D}, which will allow
studying possible interactions between 0D, 1D and 2D bosons.

\begin{figure}[htbp]
\includegraphics[width=0.5\textwidth]{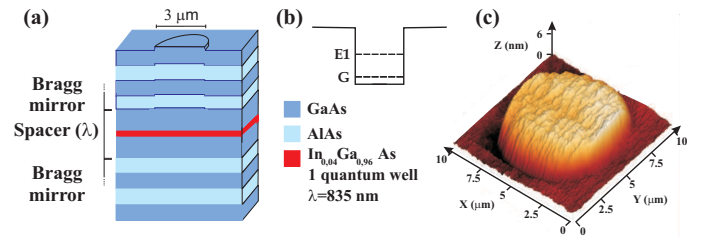}
\caption{(Color online) (a) Scheme of a circular mesa. Only the first four layers of the Bragg mirrors are represented. (b) Scheme of a potential trap
with two confined levels: the ground level (G) and one excited state (E1). (c) AFM image of a $9 \mu m$-diameter circular mesa on the surface of the top
mirror. The lateral scales are in microns and the vertical scale is in nm.} \label{mesa}
\end{figure}

\begin{figure}[htbp]
\includegraphics[width=0.5\textwidth]{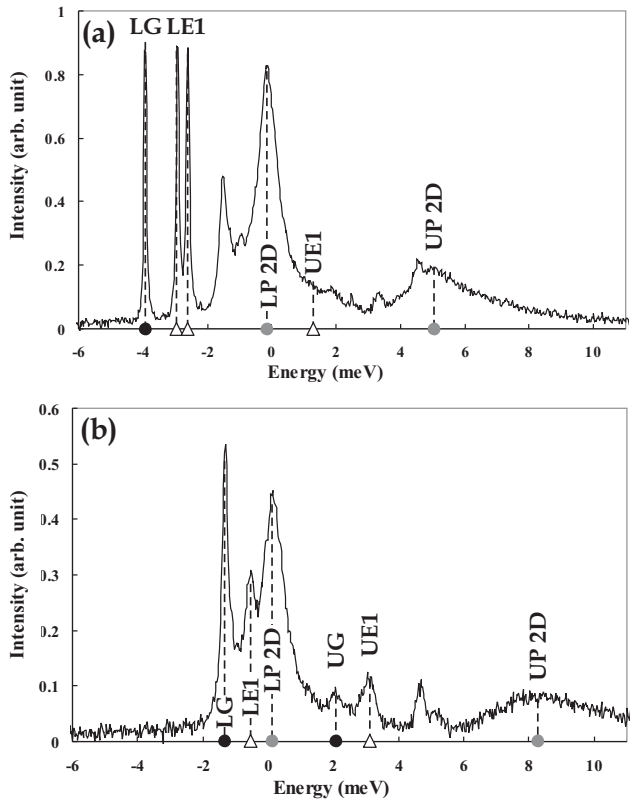}
\caption{Photoluminescence spectra of a circular $3 \mu$m diameter mesa at two different positions: (a) $x=1.04$ and (b) $x=4.42$ (see
Fig.\ref{anticrois}). The origin of the horizontal axis corresponds to the exciton's energy. Grey circles: lower (2D LP) and upper (2D UP) 2D polaritons
around the mesa. Dark circles: lower (LG) and upper (UG) polaritons of the ground level of the mesa. Triangles: lower and upper polaritons of the first
excited level (E1) of the mesa. (a) LG and LE1 polariton states are quasi-photonic. (b) The excitonic and photonic components are comparable for the
confined polariton states. Other confined levels are visible on  both spectra but are not labeled for sake of clarity.} \label{sp}
\end{figure}

By using an original technique, we were able to fabricate microcavities with mesas etched on the spacer layer. The mesa lengthens the cavity and thus
lowers the cavity mode energy (see Fig.\ref{mesa}). This local difference creates a potential trap for the confined photons. The lateral shape and size of the
mesas are defined by photolithography. The height of the mesas is determined by the etching process and can be changed at will.  It has to be large enough to allow the existence of at least one confined level. Here, it has also been chosen small enough to obtain mesas and planar cavity photon modes close to the excitonic resonance. Thus we manage to observe mixed exciton-photon states inside and outside the mesas on the same part of the sample. Such a method of defining the structure, by a step of a few nanometers only, avoids any lateral losses as observed in the micro-pillar structure previously published \cite{obert}.

The final structure, schematically represented with a circular mesa  on Fig.~\ref{mesa}(a), is prepared in the following way. First, we grow 22
AlAs/GaAs pairs of Distributed Bragg Reflectors (DBR), then the GaAs $\lambda-$spacer, with a single embedded InGaAs/GaAs quantum well (QW). The $80 \AA$
quantum well with about $4\%$ Indium in Gallium Arsenide, is chosen to allow measurements in transmission through the wafer below the gap of GaAs
\cite{HoudreAPL}. Then we perform photolithography defining the pattern of the mesas, and finally the etching, which determines their height. We can design
various patterns for the mesas on a single sample, but the etching process defines a common height for all of them. Let us stress that the excitonic mode is
absolutely not affected by the etching process, whatever the pattern of the mesa. Finally the regrowth of the upper 21 pairs is realized again by
Molecular Beam Epitaxy, after an \textit{in situ} hydrogen cleaning which guarantees a perfect regrowth interface.

The high quality of the sample can be appreciated by atomic force microscopy (AFM) measurements. Fig.~\ref{mesa}c)
shows a $9 \mu m$-diameter circular mesa on the surface of the complete structure. Amazingly enough, the $6 nm$ step height of the etched mesas is kept even after the
regrowth of the $2.5 \mu m$ thick top Bragg mirror. The abruptnesses of the steps as a function of the crystal orientation is correlated to the mobility of
the elements during growth. The profile along $[011]$ is much steeper ($0.5 \mu m$ width) than along $[0\overline{11}]$ ($3 \mu m$ width), which
corresponds to a slower mobility of the atoms. This gives rise to a surface asymmetry along the different crystalline orientations, with a smoothing of the etched
faces that increases proportionally to the overgrowth thickness. The step height corresponds to the change of the photon mode energy that we calculated by
transfer matrix simulations. According to the AFM measurement on the top mirror of a $6 nm$ step, we compute an energy difference of $9 meV$, in very good
agreement with the photoluminescence experiment results.

Let us first consider the properties of the microcavity, the effect of etching and regrowth.  For this we measure the anticrossing curve of the
resonances, resulting from the strong coupling regime of the 2D polaritons and on large square mesas of $300 \mu m$, where no confinement effect is
expected, but where the photonic resonance is red-shifted due to the larger length of the cavity. The anticrossing curve is accessible using the wedge
of about $4 \%$ in this sample. This thickness variation corresponds, for a $\lambda-$cavity, to an energy variation for the photonic resonance of about
$50 meV$ across the whole sample. In the zone of interest the variation is about $2.4 meV/mm$. As the quantum well's resonance is less affected by the
thickness, its energy variation is below $3 meV$ across the whole sample. The spectral properties are measured in a photoluminescence experiment. The
sample, cooled to about 10K, is pumped by an Argon laser ($\lambda = 532 nm$), at low pump intensity (see below). The luminescence is analyzed with a
$25\mu eV$ resolution spectrometer. The microcavity features a Rabi splitting energy of $3.5 meV$, and a full width at half the maximum (FWHM) of $220\mu
eV$ for the photon mode, corresponding to a quality factor of $Q\simeq 7*10^{3}$, and of $500 \mu eV$ for the quantum well exciton. We then
performed the same anticrossing measurement on the large square mesas of $300 \mu m$, disposed all along the wedge, separated by $300 \mu m$ one from the other.  Finally we obtained a
double anticrossing curve (not shown). The two photon modes -on the mesas and around the mesas- are separated by $9 meV$. The photon mode shows the same linewidth
on both domains. The two anticrossings have exactly the same characteristics, which guaranties that the etching process just shifts the cavity
mode, as expected.

We now focus on the effect of the quantum confinement of the polaritons in cylindrical mesas of $3 \mu m$ in diameter  (other sizes give similar results).
The laser spot diameter being $\approx 17 \mu m$, the spectra display both the confined polaritons of the mesa and the 2D polaritons of
the planar cavity outside the mesa, displaced towards higher energies. Let us note that the light emission from one single quantum box is very intense,
which allowed us to perform all experiments on single polariton boxes, well below the non-linear regime, in contrast to the case of pillar
microcavities \cite{dasbach,obert}. We consider first the photoluminescence spectrum of a mesa relatively far from the excitonic resonance on Fig.\ref{sp}a). The lower polariton mode of
the mesa is thus quasi-photonic and probes the quantum confinement of the photonic box. It is split into several peaks. For sake of clarity, we will only consider the ground level (G) and the first excited level (E1), which is a doublet. This doublet corresponds to a degeneracy lifting that results from the asymmetry of the mesas at the level of the spacer \cite{assymetry}. This asymmetry is already present on the photolithography mask and is not linked to any growth or etching process. The quantized photonic peaks have a spectral width of
about $90 \mu eV$, a factor of two smaller than observed on the non-confined cavity photon mode. The smallest value measured reached $70 \mu eV$, indicating a
quality factor of $Q\simeq 2.1*10^{4}$ for the confined photon modes (better than for the 2D photon modes, thanks to the additional lateral confinement). The energy of the ground level is $1.5 meV$ above the energy of the bottom of the trap
(as observed on the large mesas at the same position on the wedge).

\begin{figure}[htbp]
\includegraphics[width=0.5\textwidth]{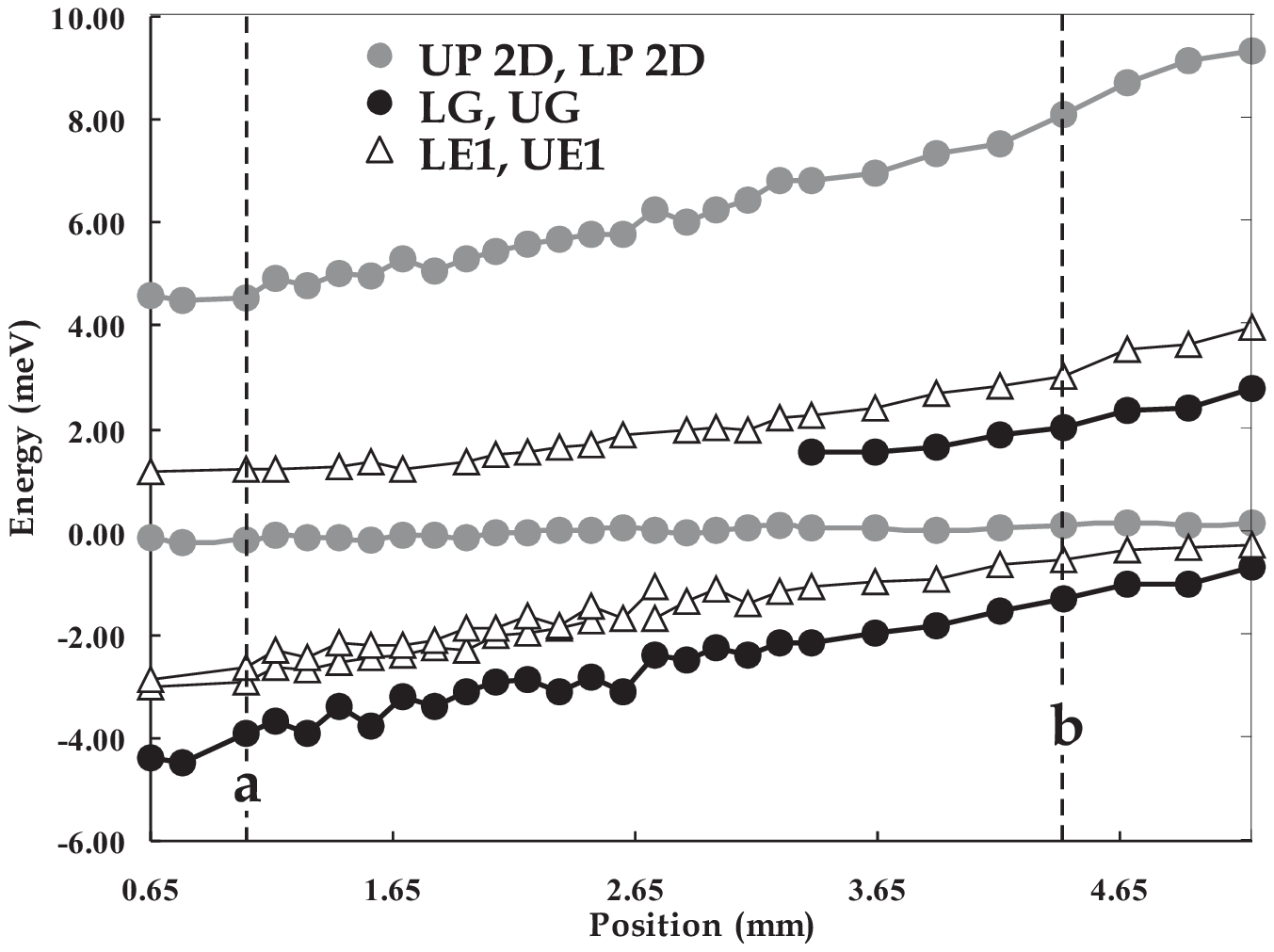}
\caption{Energy of the photoluminescence peaks as a function of the spot position along the wedge. Rabi splitting of $3.5meV$ ($3.35 meV$) for the 2D
polaritons (0D polaritons). (a) and (b) indicate the positions at which the spectra on Fig.\ref{sp} have been taken. Zero detuning for G (E1) around
$x=4.6 cm$ ($x=3.05 cm$).}. \label{anticrois}
\end{figure}

To demonstrate the strong coupling even in the smallest mesas and the polariton nature of the trapped states, we systematically measured the spectra on
mesas disposed along the wedge, i.e. with different detunings. For the 2D polaritons, we are always rather far from the position for which exciton and 2D
cavity photon are degenerate, and the lower polariton mode stays at the exciton energy on the whole range of positions shown in Fig.\ref{anticrois}. For the confined states, strong
coupling is very nicely demonstrated by the anticrossing curve for the emission of each of the two confined levels of the mesas' polaritons. They
present a Rabi splitting of around $3.35 meV$. The degeneracy lifting of the first excited level is no longer visible for the lower polariton modes when their excitonic component is too important. It is not at all visible for the upper polariton modes, due to additional relaxation channels, such as the coupling to continuum states or lower energy modes.
Fig.\ref{sp}b) displays the photoluminescence spectrum very close to zero detuning in the trap, where lower and upper polariton modes confined in the quantum box are nicely
resolved. Despite the relatively large linewidth of the 2D exciton, the high Q factor and the very efficient relaxation to the quantum boxes, allow to
observe the confinement for both the lower and upper polariton states.

In summary, we have reported the realization of a semiconductor microcavity in strong coupling regime with embedded polariton quantum boxes. The trap
consists of a mesa in the spacer layer, obtained by a well controlled technique. To our knowledge, this is the first time that quantum confinement is
resolved simultaneously for both upper and lower polariton modes, and that the anticrossing is resolved separately for all the quantized polaritons. Experimental
and theoretical work on the dispersion of the quantized modes is being performed. The co-existence in the same sample of 2D and 0D
polaritons, as well as the possibility to study a single quantum box, will allow original studies such as resonant parametric scattering or trapping by
spatial relaxation. Wide possibilities in the transverse shape of the mesas also open the way to the realization of various experimental configurations
for the bosonic polariton states, which energy, wavefunction and dimensionality can be engineered on demand.

We would like to acknowledge fruitful discussions with C. Ciuti, P. Lugan, M. Saba, and V. Savona. We are thankful for the strong financial support from the Swiss NCCR
research program Quantum Photonics.


\newpage

\begin{thebibliography}{99}

\bibitem{Weisbuch92} C. Weisbuch, M. Nishioka, A. Ishikawa et Y. Arakawa, Phys. Rev. Lett. \textbf{69}, 3314 (1992).

\bibitem{Snoke} S. A. Moskalenko and D. W. Snoke, \textit{Bose-Einstein Condensation of Excitons and Biexcitons and
Coherent Nonlinear Optics with Excitons} (Cambridge University Press, Cambridge, 2000).

\bibitem{kavokin} A. Kavokin, G. Malpuech and F.P. Laussy, Phys.Lett. A \textbf{306}, 187 (2003).

\bibitem{Baumberg00} J. J. Baumberg, P. G. Savvidis, R. M. Stevenson, A. I. Tartakovskii, M. S. Skolnick, D. M. Whittaker,and J. S. Roberts, Phys. Rev. B
\textbf{62}, R16247 (2000).

\bibitem{Richard05} M. Richard, J. Kasprzak, R. Romestain, R. Andr\'{e} and Le Si Dang Phys. Rev. Lett.  \textbf{94}, 187401 (2005).

\bibitem{Reithmaier} J. P. Reithmaier, G. Sek, A. L\"offler, C. Hoffmann, S. Kuhn, S. Reitzenstein, L. V. Keldysh, V. D. Kulakovskii, T. L. Reinecke and A. Forchel, Nature \textbf{432}, 197 (2004).

\bibitem{Yoshie} T. Yoshie, A. Scherer, J. Hendrickson, G. Khitrova, H.M. Gibbs, G. Rupper, C. Ell, O. B. Shchekin and D. G. Deppe, Nature \textbf{432}, 200 (2004).

\bibitem{Peter} E. Peter, P. Senellart, D. Martrou, A. Lema\^{\i}tre, J. Hours, J.-M. G\'{e}rard and J. Bloch, Phys. Rev. Lett. \textbf{95}, 067401 (2005).

\bibitem{bloch} J. Bloch, F. Boeuf, J.-M. G\'{e}rard, B. Legrand, J.-Y. Marzin, R. Planel, V. Thierry-Mieg, and E. Costard, Physica E \textbf{2}, 915 (1998).

\bibitem{obert} M. Obert, J. Renner, A. Forchel, G. Bacher, R. Andr\'{e}, and D. Le Si Dang, Appl. Phys. Lett \textbf{84}, 1435 (2004).

\bibitem{dasbach} G. Dasbach, M. Schwab, and A. Forchel, Phys. Rev. B \textbf{64}, 201309(R) (2001).

\bibitem{dasbach1D} G. Dasbach, A.A. Dremin, M. Bayer, V.D. Kulakovskii, N.A. Gippius, and A.Forchel, Phys. Rev. B \textbf{65}, 245316 (2002).

\bibitem{assymetry} They have an elliptical shape with an absolute difference of about  $0.75\mu m$ between the small and the large axis. This asymmetry is given by the pattern we designed on the photolithography mask.

\bibitem{HoudreAPL} R. P. Stanley, R. Houdr\'{e}, U. Osterle, M. Gailhanou, and M. Ilegems, Appl. Phys. Lett. \textbf{65}, 15 (2004).

\end{thebibliography}
\end{document}